# Kepler Mission Design, Realized Photometric Performance, and Early Science

David G. Koch[1], William J. Borucki[1], Gibor Basri[2], Natalie M. Batalha[3], Timothy M. Brown[4], Douglas Caldwell[5], Jørgen Christensen-Dalsgaard[6], William D. Cochran[7], Edna DeVore[5], Edward W. Dunham[8], Thomas N. Gautier III[9], John C. Geary[10], Ronald L. Gilliland[11], Alan Gould[12], Jon Jenkins[5], Yoji Kondo[13], David W. Latham[10], Jack J. Lissauer[1], Geoffrey Marcy[2], David Monet[14], Dimitar Sasselov[10], Alan Boss[15], Donald Brownlee[16], John Caldwell[17], Andrea K. Dupree[10], Steve B. Howell[18], Hans Kjeldsen[6], Søren Meibom[10], David Morrison[1], Tobias Owen[19], Harold Reitsema[20], Jill Tarter[5], Stephen T. Bryson[1], Jessie L. Dotson[1], Paul Gazis[5], Michael R. Haas[1], Jeffrey Kolodziejczak[21], Jason F. Rowe[1,23], Jeffrey E. Van Cleve[5], Christopher Allen[22], Hema Chandrasekaran[5], Bruce D. Clarke[5], Jie Li[5], Elisa V. Quintana[5], Peter Tenenbaum[5], Joseph D. Twicken[5], and Hayley Wu[5]

[1]NASA Ames Research Center, Moffett Field, CA 94035
[2]Dept. of Astronomy, University of California-Berkeley, Berkeley, CA 94720
[3]Dept. of Physics and Astronomy, San Jose State University, San Jose, CA 95192
[4]Las Cumbres Observatory Global Telescope, Goleta, CA 93117
[5]SETI Institute, NASA Ames Research Center, Moffett Field, CA 94035
[6]Aarhus University, DK-8000, Aarhus C, Denmark
[7]McDonald Observatory, University of Texas at Austin, Austin, TX 78712
[8]Lowell Observatory, Flagstaff, AZ 86001
[9]Jet Propulsion Laboratory/California Institute of Technology, Pasadena, CA 91109
[10]Harvard-Smithsonian Center for Astrophysics, Cambridge, MA 02138
[11]Space Telescope Science Institute, Baltimore, MD 21218
[12]Lawarence Hall of Science, Univ. California-Berkeley, Berkeley, CA 94720
[13]NASA Goddard Space Flight Center, Greenbelt, MD 20771
[14]United States Naval Observatory, Flagstaff, AZ 86002
[15]Carneige Institution of Washington, Washington, DC 20015
[16]Dept of Astronomy, University of Washington, Seattle, WA 98195
[17] Dept of Physics and Astronomy, York University, Toronto, ON M3J 1P3, Canada
[18]National Optical Astronomy Observatory, Tucson, AZ 85726
[19]Institute for Astronomy, University of Hawaii, Honolulu, HI 96822
[20]Ball Aerospace and Technologies Corp., Boulder, CO, 80306
[21] Space Science Office, NASA Marshall Space Flight Center, Huntsville, AL 35812
[22]Orbital Sciences Corp., NASA Ames Research Center, Moffett Field, CA 94035
[23]NASA Postdoctoral Program fellow,

Electronic address: D.Koch@NASA.gov

**Abstract**
The *Kepler Mission*, launched on Mar 6, 2009 was designed with the explicit capability to detect Earth-size planets in the habitable zone of solar-like stars using the transit photometry method. Results from just forty-three days of data along with ground-based follow-up observations have identified five new transiting planets with measurements of their masses, radii, and orbital periods. Many aspects of stellar astrophysics also benefit from the unique, precise, extended and nearly continuous data set for a large number and variety of stars. Early results for classical variables and eclipsing stars show great promise. To fully understand the methodology, processes and eventually the results from the mission, we present the underlying rationale that ultimately led to the flight and ground system designs used to achieve the exquisite photometric performance. As an example of the initial photometric results, we present variability measurements that can be used to distinguish dwarf stars from red giants.

Subject headings: Instrumentation: photometers – planetary systems – space vehicles: instruments – stars: statistics – stars: variables – techniques: photometric
Facilities: *Kepler Mission*

# 1. Introduction

The foremost purpose of the *Kepler Mission* is to detect Earth-size planets in the habitable zone (Kasting et al.1993) of solar-like stars (F- to K-dwarfs), determine their frequency, and identify their characteristics. The method of choice is transit photometry (Pont et al. 2009), which provides the orbital period and size of the planet relative to its star. When combined with stellar parameters and radial velocity measurements, the mass, radius and density of the planet are obtained. Transit photometry requires high photometric precision with continuous time series data of a large number of stars over an extended period of time. Although designed for the explicit purpose of terrestrial planet detection, the nature of the *Kepler* data set is also of tremendous value for stellar astrophysics. Asteroseismology (Stello et al. 2009), gyrochronology (Barnes 2003), and astrometry, the studies of solar-like stars (Basri et al. 2010, Chaplin et al. 2010), eclipsing binaries (Gimenez et al. 2006), and classical variables (Gautschy & Saio 1996) all depend on precise, extended, continuous flux time series. Results from many of these fields in turn tie back into the interpretation of exoplanet science.

*Kepler* was launched on Mar 6, 2009, commissioned in sixty-seven days and began science operations on May 13. In this issue of *Astrophysical Journal-Letters*, we present results from the first two data sets: Q0 consisting of 9.7 days of data taken during commissioning and Q1 consisting of 33.5 days of data taken before the first quarterly roll of the spacecraft. The Q0 data are particularly unique and interesting, since all spectral types and luminosity classes of stars with $V<13.6$ were included.

# 2. Fundamental requirements

Borucki et al. *(*2008) have described the scientific goals and objectives of the *Kepler Mission*. Overviews of much of the hardware have been given elsewhere (Koch et al. 2004). This section describes how the scientific goals formed the basis for the design of the mission.

*2.1 Mission duration*
The fundamental requirement for mission success is to reliably detect transits of Earth-size planets. To have confidence that the signatures are of planets, we require a sequence of at least three transits, all with a consistent period, brightness change and duration. To be able to detect three transits of a planet in the habitable zone of a solar-like star, led to the *first* requirement, a mission length of at least three years.

*2.2 Number of stars to observe*
Transit photometry requires that the orbital plane of the exoplanet must be aligned along our line-of-sight to the exoplanet's parent star. Let $\theta$ be the angle between the line-of-sight and the orbital plane for a planet that just grazes the edge of its star. Then $\sin(\theta)=R/a$, where $R$ is the sum of the radii of the star and the planet and $a$ is the semi-major axis of the orbit, taken to be nearly circular. The solid angle integrated over the sky for all orbital pole positions where the planet can be seen transiting is $4\pi\sin(\theta)$. Dividing by the total area of the sky ($4\pi$) and substituting in for $\sin(\theta)$ yields
$$\varphi=R/a \qquad (1)$$
as the geometric random probability for alignment of a transit, with no small angle approximation even down to where $a=R$. For an Earth-Sun analog, $R/a \cong 0.5\%$. The *second* requirement is to observe at least one-hundred thousand solar-like stars with ~5000 having $V\leq12$. Assuming each solar-like star has just one $R=1.0R_\oplus$ planet in or near the habitable zone, one would expect twenty-five "earths", a statistically meaningful result (Borucki et al. 2010b). A null result would also be significant.

*2.3 Photometric precision*
An Earth-Sun analog transit produces a signal of 84 parts per million (ppm), the ratio of their areas. A central transit of an Earth-Sun analog lasts for 13 hours. The *third* requirement is to reliably detect transits of 84 ppm in 6.5 hrs (half of a central transit duration for an Earth-Sun analog). There are three distinct types of noise that determine the detection threshold: 1) photon-counting shot noise, 2) stellar variability and 3) measurement noise. We define the combination of these sources as the combined differential photometric precision (CDPP).
$$\text{CDPP}=(\text{shot noise}^2 + \text{stellar variabiltiy}^2 + \text{measurement noise}^2)^{1/2}. \qquad (2)$$
Equation 2 needs to be used with caution, since stellar variability and many of the measurement noise terms do not scale simply with time. CDPP is differential, since it is the difference in brightness from the near-term trend that is important - not long-term variations. For an 84 ppm signal to be detected at $\geq 4\sigma$, the CDPP must be $\leq$20 ppm. In a well-designed experiment, little is gained by reducing any noise component to be significantly smaller than the total. We have no control over the stellar variability contribution other than selecting stars with low variability. Analysis of data from the Sun shows that variability during solar maximum on the time scale of a transit is typically 10 ppm (Jenkins, 2002). For the shot noise from a star to be 14 ppm in 6.5 hrs, the flux from the star needs to result in $5 \times 10^9$ photoelectrons. Quadratic subtraction of these two components from 20 ppm, leaves ~10 ppm for the measurement noise. Measurement noise includes not just detector and electronic noise, but also such things as pointing jitter, image drift, thermal, optical, and integration-time stability, stray light, video and optical ghosting, and sky noise.

*2.4 Stability - the key to success*
A technology demonstration was performed (Koch et al. 2000) to prove the feasibility of the *Kepler Mission*. Key lessons learned from these tests were that pointing stability had to be better than 0.003 arcsec/15 min and thermal stability of the CCD had to be better than 0.15K/day. This led to the *fourth* design requirement: photometric noise introduced by any source must either have a mean value of zero or not vary significantly on the time scale of a transit.

Several other aspects were also demonstrated: 1) the required precision can be achieved without a shutter, 2) optimum photometric apertures are necessary, 3) pixels with large full wells allow for longer times between readouts, 4) traps in the CCDs will remain filled if the CCDs are operated below about -85C and the integration times are kept <8 sec, 5) precision photometry works even with saturated pixels, if the upper parallel clock voltage is properly set, and 6) data at the individual pixel level should be preserved.

*2.5 Pixel time series*
Preserving the individual pixel time series that compose each stellar image permits the ground analysis to:
1. Measure and decorrelate the effects of image motion (Jenkins et al. 2010a),
2. Remove cosmic rays at the pixel level (Jenkins et al. 2010a),
3. Remove instrumental features and systematic noise (Caldwell et al. 2010),
4. Measure the on-orbit pixel response function (PRF) (Bryson et al. 2010),
5. Fit and remove local background for each star (Jenkins et al. 2010a),
6. Measure any centroid shift during a transit at the sub-millipixel level (Batalha et al. 2010a),
7. Perform astrometry and obtain parallaxes and hence distances to the stars (Monet et al. 2010), and
8. Reprocess the data using improved understanding of the instrument and data.

## 3. Mission design choices
*3.1 Observing strategy*
Three design considerations enter into the choice of how best to observe a large number of stars for transits: size of the field-of-view (FOV), aperture of the optics, and duty cycle. Concepts considered were large FOV fish-eye optics, multiple aperture optics, a single large aperture, and single versus multiple pointings during each cadence. After considering various combinations, the design that was settled upon was a one-meter class aperture with a FOV >100 square degrees and viewing of a single star field. Pointing to a single star field has the advantages of: 1) selecting the richest available star field, 2) minimizing the stellar classification necessary to select the desired stars, 3) optimizing the spacecraft design, 4) maximizing the duty cycle, 5) simplifying operations, data processing and data accounting, and 6) continuous asteroseismic measurements over long periods of time. The combination of these factors results in a photometric database with unprecedented precision, duration, contiguity and number and variety of stars.

For an extended mission, there are many reasons to not change the star field. Beside the points above, additional reasons are: 7) extending the maximum orbital period for planet detection, 8) improving the photometric precision for the already detected planets, 9) improving the statistics for marginal candidates, 10) decreasing the minimum detectable size of planets, 11) providing a longer time base for transit and eclipsing binary timing to enable detection of unseen companions and 12) observing stellar activity cycles that approach the time scale of a solar cycle. Based solely on the current rate of propellant usage, the only expendable, the mission can operate for nearly ten years.

*3.2 The photometer*
The Delta II (7925-10L) launch vehicle provided by NASA's Discovery program defined the combination of the size of the optics, the sunshade and solar avoidance angle, the spacecraft mass, and the orbit. In the final analysis, it was found that a one-meter class Schmidt telescope with a 16º diameter FOV and a 55º sun-avoidance angle could be launched into an Earth-trailing heliocentric orbit (ETHO). The shot noise of 14 ppm in 6.5 hr is met with the 0.95-m aperture for a $V$=12 solar-like star. The large FOV required a curved focal surface and the use of sapphire field-flattening lenses on each CCD module. The spectral bandpass is defined by the optics, the quantum efficiency of the CCDs and bandpass filters. To maximize the signal-to-noise ratio (SNR) for solar-like stars, the bandpass filters applied to the field-flattening lenses have a >5% response only from 423 to 897 nm. This is shown in figure 1. The blue cutoff was chosen to avoid the UV and the Ca II H&K lines. For the Sun, 60% of the irradiance variation is at <400nm, but photons <400nm only account for 12% of the total flux (Krivova et al. 2006). The red cutoff was selected to avoid fringing due to internal reflection of light in the CCDs. The spectral response is somewhat broader than a combination of $V$ and $R$ bands. *Kepler* magnitudes ($Kp$) are usually within 0.1 of an $R$ magnitude for nearly all stars.

We shall refer to the single instrument on the *Kepler Mission* as the photometer. The heart of the photometer is the 95-megapixel focal plane composed of 42 science CCDs and four fine guidance sensor CCDs (figure 2). The science CCDs are thinned, back-illuminated and anti-reflection coated, four-phase devices from e2v Technologies Plc. The CCD controller, located directly behind the focal plane, provides the clocking signals and digitizes the analog signals from the 84 CCD outputs. Data from each CCD are co-added on board for 270 readouts to form a long cadence (LC) of ~30 min, the primary data for planet detection. At the end of each LC, only the pixels of interest (POI) for each star are extracted from the image, compressed and stored for later downlink. In parallel, a limited set of 512 POI are extracted from the image every nine readouts, to provide short cadence (SC) 1-min data. The SC data are used to provide improved timing of planetary transits and to perform asteroseismology (Gilliland et al. 2010a). In addition to the POIs for measuring stellar brightnesses, significant amounts of collateral data are also collected to enable calibration (Caldwell et al. 2010). Full-field images (FFI) are also recorded once per month, calibrated and archived. The design and testing of the focal plane assembly is described by Argabright et al. (2008). A summary of the key characteristics is given in table 1.

*3.3 The orbit*
An ETHO is significantly more benign and stable for precision photometry than an Earth orbit. Important features of an ETHO (nearly identical to the *Spitzer* ETHO, Werner, et al. 2004) versus a low-Earth orbit are: 1) the spacecraft is not passing in and out of the radiation belts or the South Atlantic Anomaly 2) the spacecraft is not passing in and out of the Earth's shadow and heating from the Sun, 3) there is no atmospheric drag or gravity gradient to torque the spacecraft, and 4) there is no continuously varying earthshine getting into the telescope - all of these happening on the time scale of a transit. The largest disturbing torque in an ETHO is sunlight.

*3.4 Star field selection*
The 55º sun avoidance angle limited the choice to a star field >55º from the ecliptic plane. Two portions of the galactic plane are accessible. Counting stars brighter than *V*=14 from the USNO star catalog showed the Cygnus region, looking along the Orion arm of our galaxy, to be the richest choice. After reviewing the confusion due to distant background giants in the galactic plane, the center of the FOV was placed at 13.5º above the galactic plane at an *RA*=$19^h22^m40^s$ and *Dec*=44°30' 00". For distances greater than a few kiloparsecs, stars in the field are above the galactic disk, thus minimizing the number of background giant stars. The distance to a *V*=12 solar-like star is ~270 pc, with the typical distance to terrestrial planets detected by *Kepler* being 200-600 pc. An important aspect of the northern over the southern sky is that the team resources needed for ground-based follow-up observing are in the North.

The final orientation of the focal plane was chosen to minimize the number of bright stars on the CCDs. Bright stars bloom and cause significant loss of useable pixels. Only a dozen stars brighter than *V*=6 are on silicon, with only one, θ Cyg, being brighter than *V*=5. The spacecraft needs to be rotated about the optical axis every one-quarter of an orbit around the Sun to keep the Sun on the solar panels and the radiator that cools the focal plane pointed to deep space. To keep the bright stars in the gaps between the CCDs and blooming in the same direction on the CCDs, the 42 CCDs were arranged with four-fold symmetry (except for the central two CCDs) and mounted to within +/-3 pixels of co-alignment.

Given the 16º diameter FOV and the large varying angle between the center of the star field and the velocity vector of the spacecraft as it orbits the Sun, the affect of velocity aberration on the locations of the stars on the CCDs is significant. This effect causes the diameter of the FOV to change on an annual basis by 6 arcsec, and is taken into account when computing the POI used for each star.

*3.5 The Kepler Input Catalog*
To maximize the results, we preferentially observe solar-like stars. Most star catalogs provide sufficient information to approximate $T_{eff}$, but the stellar size is generally not provided. We chose two approaches to distinguish dwarf from giant stars: The primary approach was to perform multi-band photometric observations using a filter set similar to the Sloan Survey (g, r, i, z) with the addition of a filter for the Mg b line, that is especially sensitive to log(*g*), and then modeling of the observations to derive $T_{eff}$ and log(*g*). This resulted in the Kepler Input Catalog (KIC)[1]. The process of ranking and selecting stars based on multiple parameters using the KIC is described by Batalha et al. (2010b). The backup approach was to observe all the stars in the FOV brighter than *Kp*=15 early in the mission (the reason for the larger star handling capacity at the start of the mission) and, based on their measured variability, distinguish the dwarfs from giants. (This result is described below.)

## 4. Performance

*4.1 Saturation and dynamic range*
The design of the photometer called for collecting $5 \times 10^9$ photoelectrons in 6.5 hrs for a *V*=12 star. In a single 6.02 sec integration this amounts to $1.4 \times 10^6$ e⁻. For the tightest PRFs, 60% of the energy can fall into one pixel (Bryson et al. 2010), so stars brighter than *Kp*~11.5 saturate, depending on FOV location. We have set the upper parallel clock voltage on the CCDs so that the overflow electrons are preserved and fill the adjacent pixels in the same column. No electron is left behind. The photometric aperture sizes are adjusted for saturated stars and the photometric precision is preserved.

To illustrate that photometric precision is attained in saturated stars, light curves for some of the brightest stars, which are saturated by as much as a factor of one hundred, are presented in figure 3. These data also illustrate that the measurement noise must be well under 10 ppm. Stars fainter than *Kp*=15 are also useful. Noise measurements of many of these are near shot-noise limited performance, indicating the instrument is not limiting the noise. The dynamic range is at least from *Kp*=7 to *Kp*=17 (Gilliland et al. 2010b).

*4.2 The commissioning data set*
As a final step during commissioning, 9.7 days of photometric data were taken. A special set of 52,496 stars was used. This set had oversize apertures to compensate for the +/-3 pixel prelaunch uncertainty in the focal plane geometry. The geometry was determined to <0.1 pixel as part of commissioning (Haas et al. 2010). This Q0 data set included all stars brighter than *Kp*=13.6 that were fairly uncrowded, irrespective of their classification with the exception of about 160 stars with *Kp*<8.4. Bright stars known from published data to be variable or not to be dwarf stars were excluded. The primary uses of these data were to: 1) obtain an early measure of CDPP to verify the performance of the photometer, and 2) identify quiet stars that may have been either excluded from the *Kepler* target list based on their KIC classification or were not classified with confidence. Two of the planet detections being reported were unclassified stars.

*4.3 Variability of dwarf and giant stars*
From *Hubble Space Telescope* time series data it has been shown (Gilliland, 2008) that red giants are more variable than dwarfs and that the variations tend to be quasi-periodic with multiple simultaneous periods. Two control sets of 1000 stars each were formed from the Q0 data consisting of dwarfs and giants based on their KIC classification. They were selected to be bright, but not saturated (11.3≤$Kp$≤12.0), very uncrowded (very little or no light from a nearby star), and uniformly distributed over the focal plane. For the giants, $T_{eff}$ was required to be <5400K, since the radii given in the KIC is most secure for stars below this temperature. The data were not ensemble normalized or detrended beyond a fit to a cubic function. Impulsive outliers of greater than +5σ were suppressed. The modeled noise based on measured instrument performance (not including stellar variability) for stars in this narrow magnitude range is 26.5 ppm in 2 hr (14 ppm in 6.5 hrs) and almost entirely due to shot noise. The result of the analysis is shown in figure 4. The median value of CDPP for the dwarfs is less than twice the modeled noise level implying a median 2 hr stellar variability of <46 ppm, typical of the quiet Sun. An empirical fit of CoRoT data to their median noise from their equation 1 (Aigrain et al. 2009) for dwarfs at $R$=12 is 100 ppm in 2 hrs. Much, but certainly not all, of their higher noise level can be accounted for as due to a higher shot noise. The *Kepler* collecting area is 9.25 times greater than CoRoT's and the shot noise at a given magnitude is three times smaller.

---

[1] http://archive.stsci.edu/kepler/kepler_fov/search.php

The data in figure 4 show: 1) a clear distinction in variability between dwarfs and giants (>60% of the giants have variability >10 times the shot noise), 2) the observed variability of most dwarfs on the time scale of transits is small enough to permit detection of transits of Earth-size planets, and 3) the classification given in the KIC is quite reliable in distinguishing dwarfs from giants.

The data in figure 5 are the measured CDPP for all 52,496 of the stars in the Q0 data set. These data show an evident giant branch of higher variability distinct from the dwarfs. Also, the noise floor for $Kp$<13.6 is nearly coincident with the shot noise, indicating that the measurement noise is quite small. For fainter stars the ratio of the measurement noise to the smaller stellar flux becomes a significant part of the CDPP, but is still less than the shot noise.

## 5. Early results and prospects
*5.1 Exoplanets*
Results from the first forty-three days of data include detection of five transiting exoplanets (Borucki et al. 2010a, Borucki et al. 2010b, Koch et al. 2010, Dunham et al. 2010, Latham et al. 2010, Jenkins et al. 2010b), detection of the occultation and phase variation in the light curve of the previously known transiting planet HAT-P-7b (Borucki, et al. 2009, Welsh et al. 2010), and detection of hot compact transiting objects (Rowe et al. 2010). Each of the planet discovery papers presents details on one or more aspect of the different analysis and evaluation processes used for all of the planet discoveries. The precision has been found to be so good that the vetting process of candidates (Batalha et al, 2010a) has been able to use the photometric data to weed out most false positives that heretofore could not be recognized with photometry alone. For example, analysis of the photometric data for centroid motion in and out of transit is at the *one-tenth of a millipixel* level, permitting the recognition of about 70% of the candidates as false positives due to background eclipsing binaries. This led to an initial set of 21 candidates for the follow-up observing program (Gautier et al. 2010) with a low false-positive rate.

*5.2 Oscillating, pulsating, and eclipsing stars*
A full discussion of the *Kepler* early results and extensive references placing these in the context of their fields may be found in Gilliland (2010a). Measurement of the p-mode oscillations in the star HAT-P-7 has provided refined stellar parameters for this system, reducing the radius uncertainty from 10% (Pal et al., 2008) to 1% (Christensen-Dalsgaard et al. 2010). Measurement of p-modes from three G-type stars (Chaplin et al. 2010) has revealed twenty oscillation modes providing radii, masses and ages for these stars as a sample of what is to come. With a year's worth of data the depth of the convection zone can be measured. Solar-like oscillations have been measured in low-luminosity red giants (Bedding et al. 2010). The observations should be valuable for testing models of these H-shell-burning stars and shed light on the star formation rate in the local disk. Especially interesting are stars that show both eclipses and oscillations (Gilliland et al. 2010a, Hekker et al. 2010) allowing two independent means of inferring stellar properties. Over the 3.5 year baseline mission, a few thousand stars are to be observed at the 1-min cadence rate for the purposes of measuring p-mode oscillations.

Analysis of stellar pulsations leads to an understanding of the stellar interior. The main-sequence δ Sct stars with periods of about two hours and γ Dor stars with periods of about one day are particularly useful. Hybrid stars pulsating with both ranges of periods provide complementary model constraints. Previously only four hybrids were known. A search of the *Kepler* data has found 150 δ Sct and γ Dor stars and two new hybrids with the prospect for more to come (Grigahcene et al. 2010).

RR Lyrae stars like Cepheid variables follow a period-luminosity relationship, but unlike Cepheids, RR Lyrae are lower mass and luminosity and much more common. A significant fraction of RRab exhibit the still unexplained Blazhko effect with a modulation in the light curve of tens to hundreds of days (Kolenberg et al. 2010). RR Lyrae itself is in the *Kepler* FOV and is being observed. A search for variability in a set of 2288 stars (the bulk of which are red giants by design) has identified many new variables: 27 RR Lyrae subtype ab, 28 β Cep and δ Sct, 28 slowly-pulsating Bs and γ Dor, 23 ellipsoidal variables and 101 eclipsing binaries (Blomme et al. 2010).

*5.3 Anticipated results*
Distances to the stars can be derived from trigonometric parallax, which along with their luminosities provides the sizes of the stars, essential for knowing the size of each planet. Using PRF fitting for astrometry requires good photometry. From Q1 data it can be seen that the *Kepler* measurements are in a new regime of high SNR for astrometry (Monet et al. 2010). Astrometric precision of individual 30-min cadences is already known to be at the millipixel level (4 milliarcsec). The ultimate precision of such high SNR data is yet to be realized. Parallaxes and proper motions require years of data to separate the two.

Knowing the mass of a star is essential for calculating the orbit of a planet from the period of the transits and for calculating the mass of the planet from radial velocity observations. As solar-like stars age their rotation rates decrease (Skumanich 1972). For young clusters like the Hyades a mass-rotation-rate relationship has been derived (Barnes, 2003). *Kepler* observations of clusters NGC 6866 (0.5 Gyr), 6811 (1 Gyr), 6819 (2.5 Gyr), and 6791 (8-12 Gyr) will provide rotation rates of stars in much older clusters. From ground-based spectroscopic observations and asteroseismic analysis of the *Kepler* data (Stello et al. 2010) the masses will be derived and a mass-age-rotation rate (gyrochronology) calibration can be derived.

## 6. Summary

The *Kepler Mission* design was based on a top-down scientific requirements flow to optimize the ability to detect Earth-size planets in the habitable zone of solar-like stars. The photometer and spacecraft were launched on March 6, 2009, commissioned in 67 days and have been gathering exquisite photometric data. Analysis of the initial data has shown:

1. The discovery of five transiting planets with measured radii, masses and orbits;
2. A wealth of results from oscillating, pulsating, and eclipsing stars;
3. High photometric precision is being achieved, even for extremely saturated stars;
4. The median variability for dwarf stars is similar to that of the Sun, with many examples of stars even quieter than the Sun; and
5. Dwarf and red-giant stars can readily be distinguished based on their inherent variability.

The early data from *Kepler* have produced both new planet detections and new results in stellar astrophysics, which bodes well for future prospects.

## Acknowledgements

*Kepler* was selected as the tenth Discovery mission. Funding for this mission is provided by NASA's Science Mission Directorate. The *Kepler* science team cannot possibly name all of those who contributed to the design, development, and operation of the mission starting as far back as 1992, but we are nonetheless deeply indebted to and thankful to those hundreds of individuals. Managers at various times and institutions included: Charlie Sobeck, Marcie Smith, Dave Pletcher, Sally Cahill, Roger Hunter, Laura MacArthur-Hines, Dave Mayer, Mark Messersmith, Janice Voss, Larry Webster, John Troeltzsch, Jerome Stober, Alan Frohbieter, Monte Henderson, Len Andreozzi, Tim Kelly, Scott Tennant, Jim Fanson, Peg Frerking, Leslie Livesay, Chet Sasaki, Bill Possel and Dave Taylor; System engineers at various times and institutions included: Eric Bachtell, Dave Acton, Jeff Baltrush, Adam Harvey, Sean Lev-Tov, Dan Peters, Vic Argabright, Riley Duren, Brian Cook, Tracy Drain, Karen Dragon, Don Eagles, Rick Thompson, Steve Jara, Chris Middour, Rob Nevitt, Jeneen Sommers, Brett Strooza, Steve Walker, and Daryl Swade; and IPT leads at Ball Aerospace included: Bruce Bieber, Bill Bensler, Dan Berry, Susan Borutzki, Chris Burno, Brian Carter, Wayne Davis, Mike Dean, Bill Deininger, Mike Hale, Roger Lapthorne, Scot McArthur, Chris Miller, Rob Philbrick, Amy Puls, Charlie Schira, Dan Shafer, Beth Sholes, Chris Stewart, Dennis Teusch, Bryce Unruh, and Mike Weiss.

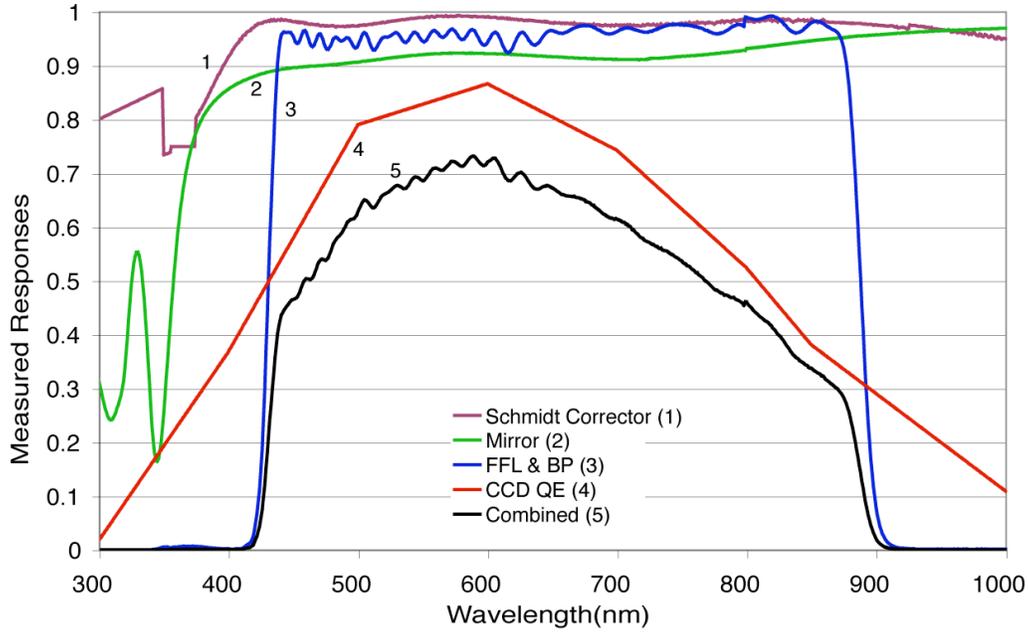

Figure 1. Spectral response curves. An average value is shown for the 21 field-flattening lenses (FFL) and the bandpass (BP) filters, which are multi-layer coatings on the back of the sapphire FFLs. The CCD quantum efficiency (QE) is the average from all 42 CCDs. The combined response curve also includes a 4% loss for contamination

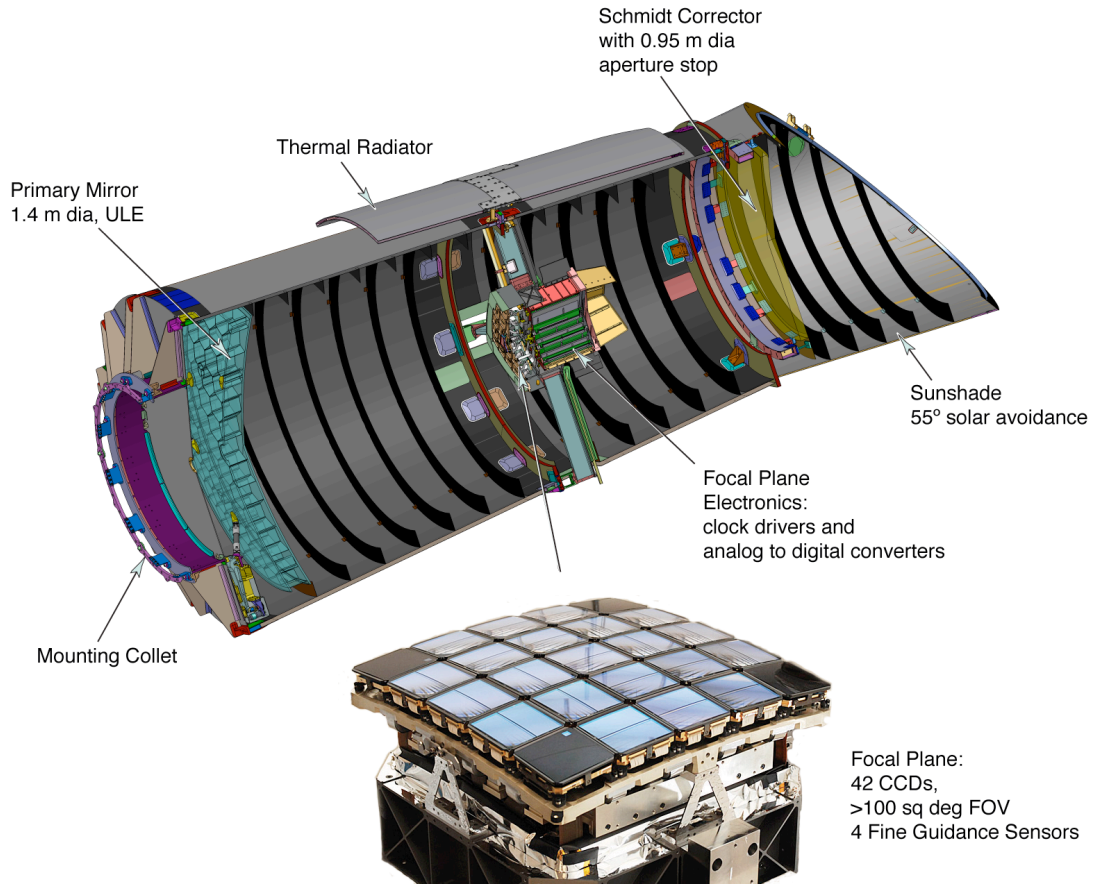

Figure 2 Cross-sectional view of the photometer and inset of the focal plane. The overall height with the sunshade is 4.3 m. The spacecraft (not shown) is a hexagonal structure 0.8 m high that surrounds the base of the photometer.

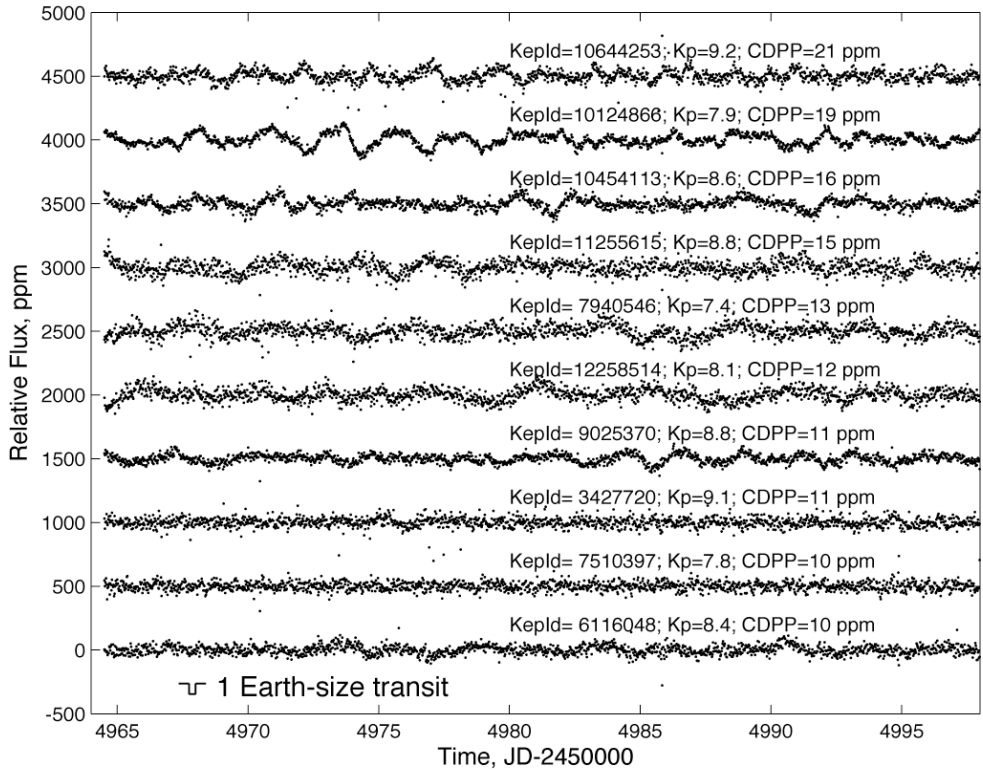

Figure 3 Flux time series for ten of the brighter G-dwarf stars. The CDPP is calculated by applying a moving-median filter that is 48 hours wide and then computing the RMS deviation for 6.5-hours. For reference, a grazing one-Earth-size transit (84 ppm) is shown to scale.

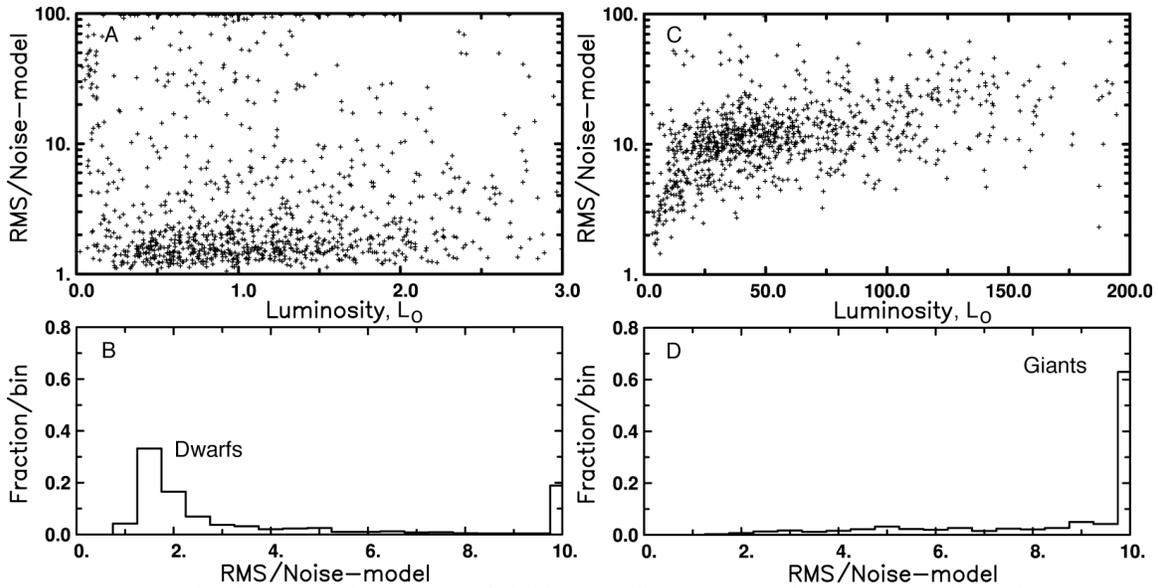

Figure 4 Distribution functions of variability for flux time series of stars classified as dwarfs and giants. Panels A & C show the ratio of RMS noise to model noise as a function of luminosity. Panels B & D show histograms for these ratios. The right hand bins include any noise ratios >10. Nineteen percent of the dwarfs in panel B are in the >10 bin. Of these 87% are recognized as intrinsic variable stars from their light curves and the remaining 13% have light curves characteristic of red giants, suggesting that the error in the classification of dwarfs in the KIC is of the order of 2.5%. Inspection of the light curves for the red giants does not show any stars that are likely to be dwarfs, based on the character of their variability.

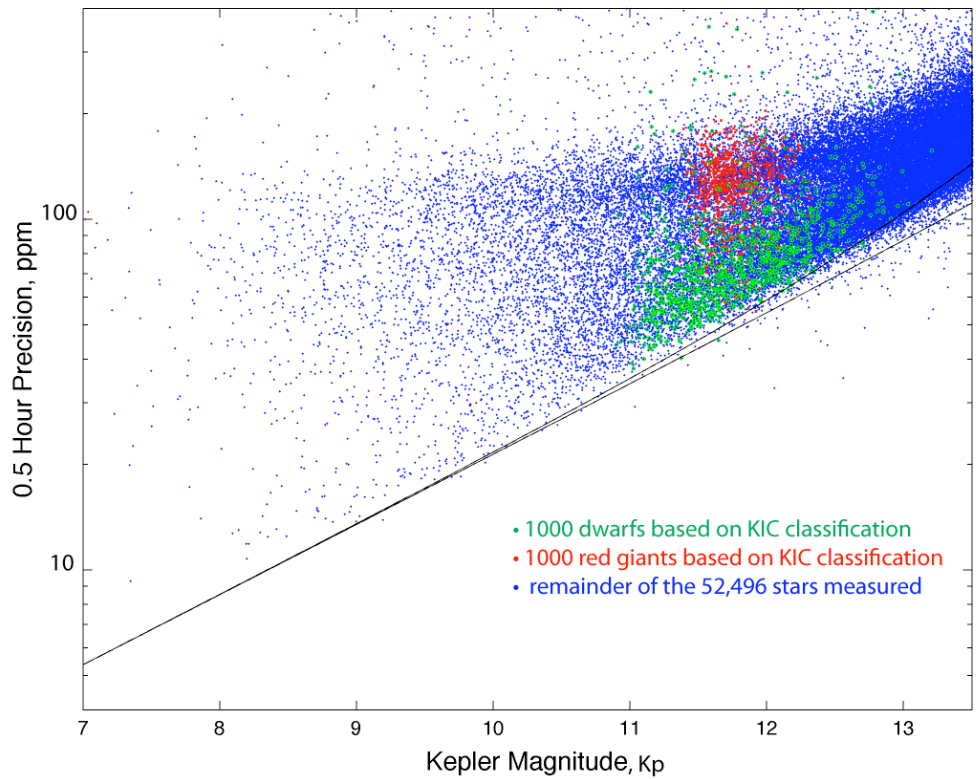

Figure 5. Measured half-hour precision of the Q0 data set. A strong separation in photometric variability can be seen between the dwarfs (green points) and the red giants (red points). The two lines bound the upper and lower noise uncertainties propagated through the data processing pipeline (Jenkins, et al, 2010c), which is mostly shot noise for $Kp<13$.

**Table 1**

*Kepler Mission* Characteristics

| Component | Value | Comment |
|---|---|---|
| **Optics** | Brashear & Tinsley | Manufacturers |
| Schmidt Corrector | 0.99 m dia./ 0.95 m dia. stop | Corning Fused silica |
| Primary mirror | F1 1.40 m dia., silver coated | Corning ULE® |
| Central obscuration | 23.03% | Due to focal plane and spider |
| Field flattening lenses | 2.5° square | Sapphire |
| Pixel Response Function | 3.14-7.54 pixels, 95% encircled energy diameter | Depends on FOV location |
| Sunshade | 55º sun avoidance | From center of FOV |
| **CCDs** | e2v Technologies | Manufacturer |
| Format | 1024 rows x 2200 columns | 2 outputs per CCD |
| Pixel size | 27 μm square | Four phase |
| Plate scale | 3.98 arcsec/pixel | |
| Full well | $1.05 \times 10^6$ electrons, typical | Set by parallel clock voltage |
| Dynamic range | $7 \leq Kp \leq 17$ | Meets photometric precision |
| Operating temp | -85C | 10mK stability |
| **Controller** | Ball Aerospace | Design and manufacturer |
| Channels | 84 | Multiplexed into 20 ADCs on 5 electronic board pairs |
| CCD integration time | 6.02 sec | Selectable 2.5 to 8 sec |
| CCD readout time | 0.52 sec | Fixed |
| Long cadence period | 1765.80 sec | 270 integrations + reads |
| Short cadence period | 58.86 sec | 9 integrations + reads |
| Maximum LC targets | 170,000 | Average 32 pixels/target |
| Maximum SC targets | 512 | Average 85 pixels/target |
| Timing accuracy | 50 msec | For asteroseismology and transit timing |
| **System** | Ball Aerospace | Design, integration, and test |
| Spectral response | 423 to 897 nm | 5% points |
| FOV | 105 sq. degrees <10% vignetted | 115 sq. degrees of non-contiguous active silicon |
| Pointing jitter | 3 milliarcsec per 15 min | 1σ per axis |
| CDPP (total noise) | 20 ppm in 6.5 hr for >90% of FOV | V=12 solar-like star including 10 ppm for stellar variability |
| Data downlink period | 31 day average | <1 day observing gap |
| Mission length | 3.5 years | Baseline. May be extended |
| Orbital period | 372 days by year 3.5 | Earth-trailing heliocentric |